\documentclass[12pt]{article}

\usepackage{graphics,graphicx,fullpage,natbib,multirow}
\usepackage{amsmath,amssymb,verbatim}
\usepackage[dvipsnames,usenames]{color}
\usepackage{caption,subcaption,booktabs}

\newtheorem{defin}{\bf Definition}

\def\ga{\mbox{Ga}}

\def\po{\mbox{Po}}

\def\E{\mbox{E}}
\def\V{\mbox{Var}}

\def\Cr{\mbox{Corr}}
\def\Cv{\mbox{Cov}}

\def\rest{\mbox{rest}}

\def\bx{{\bf x}}

\def\bz{{\bf z}}

\def\bX{{\bf X}}

\def\bZ{{\bf Z}}

\newcommand{\btheta}{\boldsymbol{\theta}}

\newcommand{\DD}{\mathcal{D}}

\begin{document}

\baselineskip=24pt

\title{\bf Modelling dependence within and across run-off triangles for claims reserving}
\author{{\sc Luis E. Nieto-Barajas$^1$ and Rodrigo S. Targino$^2$} \\[2mm]
{\sl $^1$Department of Statistics, ITAM, Mexico} \\[2mm]
{\sl $^2$School of Applied Mathematics, FGV, Brazil} \\[2mm]
{\small {\tt lnieto@itam.mx} and {\tt rodrigo.targino@fgv.br}} \\}
\date{}
\maketitle

\begin{abstract}
We propose a stochastic model for claims reserving that captures dependence along development years within a single triangle. This dependence is of autoregressive form of order $p$ and is achieved through the use of latent variables. We carry out bayesian inference on model parameters and borrow strength across several triangles, coming from different lines of businesses or companies, through the use of hierarchical priors. 
\end{abstract}

\vspace{0.2in} \noindent {\sl Keywords}: Autoregressive processes, gamma process, IBNR, latent variables.

\section{Introduction}
\label{sec:intro}

One of the largest liabilities of an insurance company is its future claims, therefore estimation of adequate reserves for outstanding claims is one of the main activities of actuaries in insurance and a major topic in actuarial science. The need to estimate future claims, given available information about the past, has led to the development of many loss reserving models. 

In this paper we study the problem of claims reserving using several run-off triangles, each of them coming from different lines of business or from different companies. For each triangle we propose a stochastic dependence model of autoregressive form of order $p$. We pull strength across lines of business/companies by considering a hierarchical prior under a Bayesian inferential approach. 

In the sequel we discuss a strand of the claims reserving literature related to this work. For an in-depth introduction to the topic the reader is referred to \cite{taylor2012loss} or \cite{wuthrich2019non}.

The oldest and most widely used technique for reserving is the ``chain-ladder'' (CL), which, due to its widespread utilization and ease of implementation, is frequently taken as a benchmark. Although the CL has been originally derived as a purely deterministic algorithm, several stochastic models provide the same predictors. For instance, \cite{mack1993distribution} provides a simple ``distribution free'' CL model, using just moment assumptions. One of the first models with distributional assumptions on the claims payments was the over-dispersed Poisson (ODP) model, proposed by \cite{renshaw1998stochastic} and popularized by \cite{england2002stochastic}. Among several other techniques, the ODP model is implemented in the \texttt{R} package \texttt{ChainLadder} \citep{ChainLadder}.

Early Bayesian stochastic formulations of the reserving problem for one line of business can be found in \cite{verrall:91}, \cite{dealba:02} and \cite{ntzoufras:02}. A Bayesian formulation of the CL model is provided in \cite{gisler2008credibility}. More recent examples include \cite{antonio2008issues}, \cite{dealba&nieto:08}, \cite{peters2009model}, \cite{meyers2009stochastic}, the monograph \cite{meyers2015stochastic}, the survey paper \cite{taylor2015bayesian}, \cite{gao2018stochastic} and the recent book \cite{gao2018bayesian}. Going down a different route and interpreting the run-off triangles as a spatially-organized data set \cite{lally2018estimating} use Gaussian process regression techniques to estimate the reserves.

On the other hand, \cite{pinheiro:03} showed that in some cases the gamma model presents a better fit to the observed values than other models that are used more frequently. Under the Bayesian paradigm, the assumption of conditionally gamma-distributed claims is made by several authors. For example, \cite{dealba&nieto:08} introduced a correlated gamma model with order of dependence one; \cite{gisler2006estimation} and \cite{merz2015claims} propose similar Bayesian gamma models, differing mainly by their choices of priors; \cite{peters2017full} study a similar model, but differently from \cite{gisler2006estimation} and \cite{merz2015claims} estimated the parameters under the Bayesian paradigm.

Over the past few years several authors proposed Bayesian models for multivariate loss reserving. One of the first contributions to this literature is \cite{merz2010paid}, where the authors develop a log-normal paid-incurred chain (PIC) model. This model is further extended in \cite{peters2014copula}. \cite{shi2012bayesian} (see also the discussion in \cite{wuthrich2012bayesian}) use a multivariate log-normal model for incremental claims to examine calendar year effects when multiple triangles are available, while \cite{merz2013dependence} assumes a log-normal model for the log-link ratios. \cite{zhang2013predicting} performs full Bayesian inference for models defined through several combinations of copulas and marginal distributions and \cite{avanzi2016stochastic} explores a multivariate Tweedie family of models. \cite{peters2017bayesian} and \cite[Chapter 6]{gao2018bayesian} study the inferential procedure of models with multiple triangles coupled through a copula. The first performs Bayesian inference for the marginals and assume the dependence structure is fully determined by the regulator. The latter compares both the inference for marginals estimation (IFME), which performs Bayesian inference for the marginals and maximum likelihood for the copula parameters, with the fully Bayesian estimation, finding similar results with less computational time with the latter.


Similarly to our proposal, other papers, such as \cite{guszcza2008hierarchical}, \cite{zhang2012bayesian}, \cite{shi2016credibility} and \cite{shi2017multivariate}, also introduce correlation across accident years through the use of hierarchical models.

To formally state the problem we denote by $X_{i,j,k}$ either the incremental claim amounts or the number of claims arising from year of origin $i$ and paid in development year $j$ for business $k$ (which can be either different lines of business within the same company, or the same line of business across different companies). Let us assume that for each $k$ we are in calendar year $n$ and the available information is given by $\mathcal{D}_n = \bigcup_{k=1}^K \DD_n^k$, where
\[ \DD_n^k = \left\{ X_{i,j,k} \, : \, i=1,\ldots,n, \ j=1,\ldots, n-i+1 \right\}. \] These available data can be represented in terms of a run-off triangle as the one given in Table \ref{tab:rotk}. 

The problem consists in predicting the quantities $X_{i,j,k}$, for $i=2,3,\ldots,n$ and $j=n+2-i,n+3-i,\ldots,n$, which correspond to the right-lower triangle in Table \ref{tab:rotk}. In particular, more than predicting individual values $X_{i,j,k}$, we are interested in the prediction of outstanding claims and thus having the necessary information to constitute adequate reserves. Let $R_{i,k}=\sum_{j=n+2-i}^{n}X_{i,j,k}$ the total aggregate outstanding claims for each year of origin $i=2,\ldots,n$ and business $k$. Moreover, the total outstanding claims for business $k$, considering all years, is $R_k=\sum_{i=2}^n R_{i,k}$, and the grand total becomes $R=\sum_{k=1}^K R_k$. 

Let us assume that the probabilistic model for $X_{i,j,k}$ is described conditional on some parameter vector $\btheta$. Then, for business $k$, all information on the outstanding reserves at time $n$ is given by the distribution of $R_{i,k} \, | \, \DD_n$, 
\[ p (R_{i,k} \mid \DD_n)  =  \int p( R_{i,k} \mid \btheta, \DD_n) p(\btheta \mid \DD_n) d\btheta, \]
which is written as an average of the model $p( R_{i,k} \mid \btheta, \DD_n)$ over all possible parameters, weighted by their posterior probability, $p(\btheta \mid \DD_n)$.

Before proceeding we introduce notation: $\ga(\alpha,\beta)$ denotes a gamma density with mean $\alpha/\beta$ and variance $\alpha/\beta^2$; $\po(\gamma)$ denotes a Poisson density with mean and variance $\gamma$.

\section{Dependent gamma model}
\label{sec:model}

Let $\{X_{i,j,k}\}$ be the set of variables of interest, for origin year $i=1,\ldots,n$, development year $j=1,\ldots,n$, and business $k=1,\ldots,K$. For a particular claim $X_{i,j,k}$, we propose a dependence model of autoregressive nature for the claims made in the previous $p$ development years, say $j-1,\ldots,j-p$. To achieve our objective we define a set of latent variables $\bZ=\{Z_{i,j,k}\}$, one for each $(i,j,k)$. 

Therefore, our model is defined through a two-level hierarchical specification of the form 
\begin{align}
\nonumber
X_{i,j,k}\mid\bZ\sim&\;\ga\left(\alpha_{i,k}+\sum_{l=0}^p Z_{i,j-l,k}\,,\,\beta_{j,k}+\sum_{l=0}^p\gamma_{j-l,k}\right),\\
\label{eq:mod}
Z_{i,j,k}\sim&\;\po\left(\alpha_{i,k}\gamma_{j,k}\right)
\end{align}
independently for $i,j=1,\ldots,n$ and $k=1,\ldots,K$, where $\{\alpha_{i,k}\}$, $\{\beta_{j,k}\}$ and $\{\gamma_{j,k}\}$ are all nonnegative parameters, and $p\geq 0$ is the order of dependence across development years. We define $Z_{i,j,k}=0$ with probability one (w.p.1), and $\gamma_{j,k}\equiv 0$, for $j\leq 0$. We will refer to \eqref{eq:mod} as dependent gamma model (DGM). 

The role of the latent variables $\bZ$ is to introduce dependence across development years. If $\gamma_{j,k}=0$ then $Z_{i,j,k}=0$ w.p.1 so the influence (dependence) of that specific development year $j$ with future years is null. If $\gamma_{j,k}=0$ for all $j$ and $k$, the $X_{i,j,k}$ become independent. Alternatively, independence across all development years can be achieved by taking $p=0$. 

The choice of the Poisson distributions for the latent variables is convenient to obtain an appealing interpretation of the model. By taking iterative expectations in model \eqref{eq:mod}, the marginal expected value of each $X_{i,j,k}$ becomes
\begin{equation}
\label{eq:mean}
\mu_{i,j,k}=\E\left(X_{i,j,k}\right)=\frac{\alpha_{i,k}\left(1+\sum_{l=0}^p\gamma_{j-l,k}\right)}{\beta_{j,k}+\sum_{r=0}^p\gamma_{j-r,k}}=\alpha_{i,k}\pi_{j,k},
\end{equation}
where $\pi_{j,k}=\left\{1+\sum_{l=0}^p\gamma_{j-l,k}\right\}/\left\{\beta_{j,k}+\sum_{r=0}^p\gamma_{j-r,k}\right\}$
are development year specific weights.
When we take $\gamma_{j,k}=0$ for all $j$ and $k$, these weights become $\pi_{j,k}=1/\beta_{j,k}$ so $\mu_{i,j,k}=\alpha_{i,k}/\beta_{j,k}$. 

Since $\sum_{i=1}^n\pi_{j,k}\neq 1$, we propose the following transformations to define interpretable quantities:
\begin{equation}
\label{eq:estim}
\alpha_{i,k}^*=\alpha_{i,k}\sum_{j=1}^n\pi_{j,k}\quad\mbox{and}\quad\pi_{j,k}^*=\frac{\pi_{j,k}}{\sum_{l=1}^n\pi_{l,k}}
\end{equation}
so that equation \eqref{eq:mean} can be written as $\mu_{i,j,k}=\alpha_{i,k}^*\pi_{j,k}^*$, where $\sum_{j=1}^n\pi_{j,k}^*=1$. In this case $\alpha_{i,k}^*=\sum_{j=1}^n\mu_{i,j,k}$ can be interpreted as the ultimate total amount for business $k$ at origination year $i$, and  $\pi_{j,k}^*$ can be interpreted as the proportion of $\alpha_{i,k}^*$ corresponding to development year $j$ in business $k$. 

The marginal variance for each $X_{i,j,k}$ can also be computed by using iterative variance results. This has the form, 
\begin{equation}
\label{eq:var}
\V\left(X_{i,j,k}\right)=\frac{\alpha_{i,k}\left(1+2\sum_{l=0}^p\gamma_{j-l,k}\right)}{\left(\beta_{j,k}+\sum_{r=0}^p\gamma_{j-r,k}\right)^2}.
\end{equation}

Additionally, as a measure of the dependence induced by our model, we can compute in closed form the covariance between any two claims for development years $j$ and $j+s$, with $1\leq s\leq p$, for the same origin year $i$ and the same business $k$. This becomes
\begin{equation}
\label{eq:covar}
\Cv(X_{i,j,k},X_{i,j+s,k})=\frac{\alpha_{i,k}\sum_{l=0}^{p-s}\gamma_{j-l,k}}{\left(\beta_{j,k}+\sum_{l=0}^p\gamma_{j-l,k}\right)\left(\beta_{j+s,k}+\sum_{l=0}^p\gamma_{j+s-l,k}\right)}
\end{equation}
and takes the value of zero for $s>p$ or if the two claims come from different origin years ($i's$) or different business ($k's$). 

Finally, with expressions \eqref{eq:var} and \eqref{eq:covar}, we can easily compute the correlation between $X_{i,j,k}$ and $X_{i,j+s,k}$, for $1\leq s\leq p$, which has the form
\begin{equation}
\label{eq:cor}
\Cr(X_{i,j,k},X_{i,j+s,k})=\frac{\sum_{l=0}^{p-s}\gamma_{j-l,k}}{\sqrt{1+2\sum_{l=0}^{p}\gamma_{j-l,k}}\sqrt{1+2\sum_{l=0}^{p}\gamma_{j+s-l,k}}}.
\end{equation} 
We note that expression \eqref{eq:cor} does not depend on the parameters $\alpha_{i,k}$ and $\beta_{j,k}$, it is only a function of parameters $\{\gamma_{j,k}\}$. Since expression \eqref{eq:cor} does not depend on the origin year $i$, we will denote the correlation as $\rho_{j,j+s,k}$ for $s\leq p$. Specifically, the numerator is a function of the parameters shared by development years $j$ and $j+s$, that is, $\gamma_{l,j}$ for $l=j+s-p,j+s-p+1,\ldots,j$. Larger/smaller values of $\gamma_{l,k}$ induce a larger/smaller correlation. 

Therefore, the set $\{\gamma_{j,k}\}$ controls de degree of dependence across developments years in the business specific triangle $k$, for instance, if $\gamma_{j,k}=0$ for all development years $j$, the correlation becomes zero between any two claims for that specific business $k$. In general, for values of $\gamma_{j,k}>0$, the dependence (in terms of correlation), induced by our model \eqref{eq:mod}, is positive and takes values in the whole range, i.e. $\rho_{j,j+s,k}\in[0,1]$, which is useful in the modelling of trends along development years in a run-off triangle.

\section{Bayesian inference}
\label{sec:inference}

Recall that available data consists of run-off triangles as those depicted in Figure \ref{tab:rotk}, that is, $\bX=\{X_{i,j,k}\}$ for $i=1,\ldots,n$, $j=1,\ldots,n+1-i$ and $k=1,\ldots,K$. To carry out a full Bayesian analysis of the model we rely on data augmentation techniques \citep[e.g.][]{tanner:91} to deal with the unobserved latent variables $\bZ$. If we denote by $\btheta=\{\alpha_{i,k},\beta_{j,k},\gamma_{j,k}\}$ the set of all model parameters, the likelihood, assuming that we have observed the latent variables, is simply the joint distribution of the observed data as well as the latent variables, that is,
$$f(\bx,\bz\mid\btheta)=\prod_{i=1}^n\prod_{j=1}^{n+1-i}\prod_{k=1}^K\{\ga(x_{i,j,k}\mid\alpha_{i,k}+\sum_{l=0}^p z_{i,j-l,k}\,,\,\beta_{j,k}+\sum_{l=0}^p\gamma_{j-l,k})$$ $$\hspace{2cm}\times\po\left(z_{i,j,k}\mid\alpha_{i,k}\gamma_{j,k}\right)\}.$$

To borrow strength across different triangles in the estimation procedure, we propose a hierarchical prior of the form 
\begin{align}
\nonumber
\alpha_{i,k}\mid a_{\alpha i},b_{\alpha i}\sim\ga(a_{\alpha i},b_{\alpha i})\quad\mbox{with}\quad a_{\alpha i}\sim\ga(a_{\alpha 0},b_{\alpha 0})\quad\mbox{and}\quad b_{\alpha i}\sim\ga(a_{\alpha 0},b_{\alpha 0}) \\
\label{eq:prior}
\beta_{j,k}\mid a_{\beta j},b_{\beta j}\sim\ga(a_{\beta j},b_{\beta j})\quad\mbox{with}\quad a_{\beta j}\sim\ga(a_{\beta 0},b_{\beta 0})\quad\mbox{and}\quad b_{\beta j}\sim\ga(a_{\beta 0},b_{\beta 0}) \\
\nonumber
\gamma_{j,k}\mid a_{\gamma j},b_{\gamma j}\sim\ga(a_{\gamma j},b_{\gamma j})\quad\mbox{with}\quad a_{\gamma j}\sim\ga(a_{\gamma 0},b_{\gamma 0})\quad\mbox{and}\quad b_{\gamma j}\sim\ga(a_{\gamma 0},b_{\gamma 0}),
\end{align}
conditionally independent for $k=1,\ldots,K$, with suitable values $a_{\alpha 0}$, $b_{\alpha0}$, $a_{\beta0}$, $b_{\beta0}$, $a_{\gamma0}$ and $b_{\gamma 0}$ to allow for low or high borrowing of strength across $k$.

Posterior distribution is characterised through the full conditional distributions of all model parameters $\btheta$ plus the conditional distributions for the latent variables $\bZ$, as well as the conditional distributions of the hyper-parameters of the hierarchical prior specification. These are given in the Appendix. Posterior inference therefore requires the implementation of a Gibbs sampler \citep{smith&roberts:93}. Since all distributions involved in our model are of standard form, MCMC procedures can also be implemented in the \texttt{R} package \texttt{rjags} \citep{rjags}. 

\section{Results}
\label{sec:results}

\subsection{Simulated data}

Before presenting a detailed application of the proposed DGM on a real data set, we briefly discuss its performance on a simulation study.

For this example we fix the number of lines of business $K=2$ and assume the claims are fully developed after $n=4$ years. We then generate a data set from the model described in (\ref{eq:mod}), with the following specifications: $p=1$; $\alpha_{i,1} = 1$ and $\alpha_{i,2} = 2$ for $i=1,\ldots,4$; $\beta_{j,k}=1$ $\forall j,k$; and $\{\gamma_{j,k},\,j=1,\ldots,4\} = \{1,4,6,2\}$, for $k=1,2$. In this case $\alpha_{i,1}^*=4$ and $\alpha_{i,2}^*=8$ for $i=1,\ldots,4$; and $\pi^*_{j,k}=1/4$ $\forall j,k$. Additionally, $\rho_{1,2,k}=0.174$, $\rho_{2,3,k}=0.263$ and $\rho_{3,4,k}=0.317$, for $k=1,2$. Since we are only changing the value of $\alpha_{i,k}$ for $k=1,2$, the proportions and correlations across development years remain the same in both triangles. 

To perform inference on this model we run two independent Markov chains, both with priors as discussed in Section \ref{sec:inference} and hyperparameters $a_{\alpha 0} = 2$, $b_{\alpha 0} = 1$, $a_{\beta 0} = 2$, $b_{\beta 0} = 2$, and $a_{\gamma 0} = 3$, $b_{\gamma 0} = 1$. The burn-in was set to $10,000$ samples and $10,000$ samples were kept for each chain.

\begin{figure}
	\centering
	\begin{subfigure}[b]{0.7\textwidth}
		\includegraphics[width=\columnwidth, page=1]{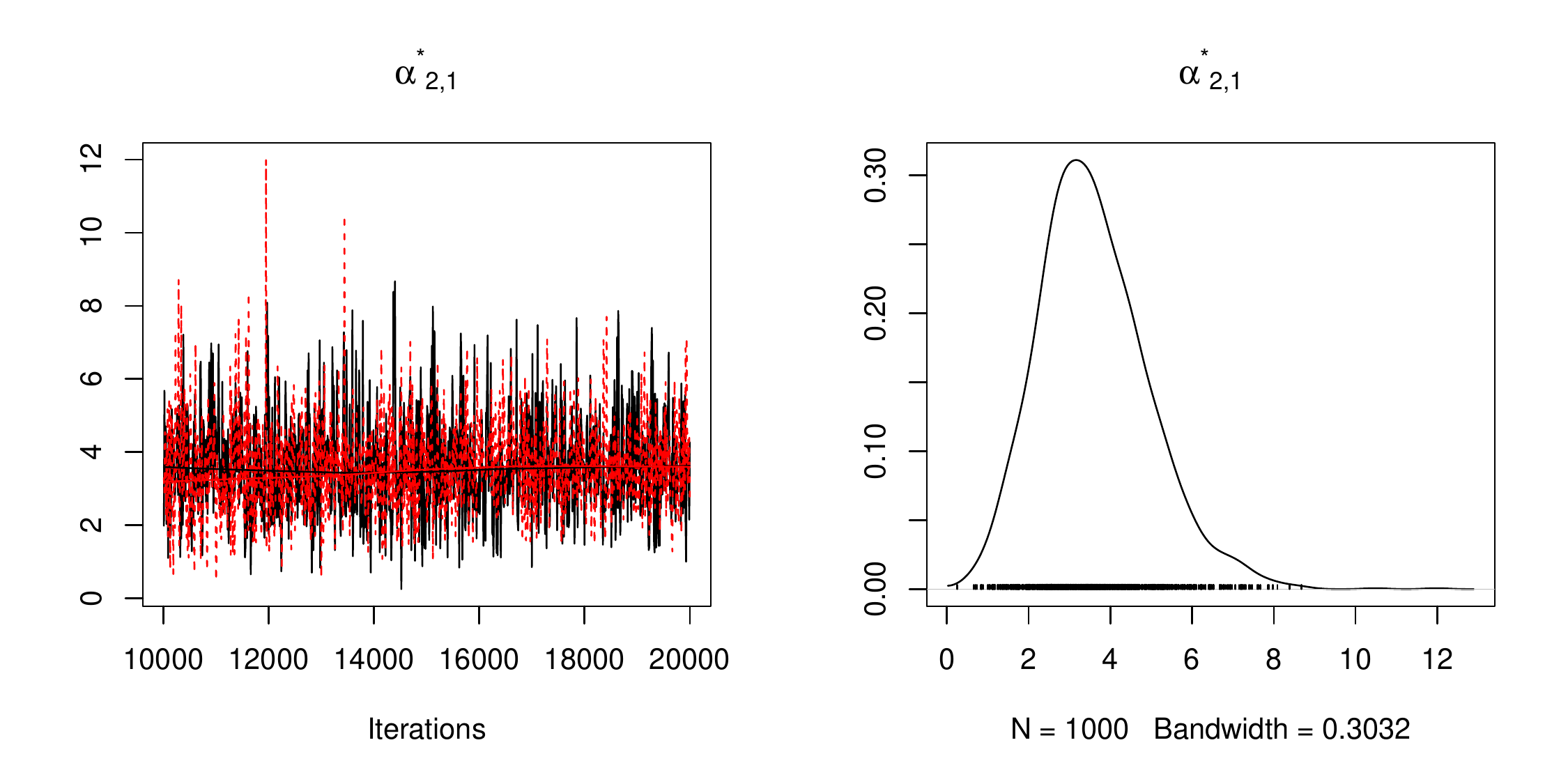}
	\end{subfigure}
	
	\begin{subfigure}[b]{0.7\textwidth}
		\includegraphics[width=\columnwidth, page=2]{2_some_trace_plots.pdf}
	\end{subfigure}

	\begin{subfigure}[b]{0.7\textwidth}
	\includegraphics[width=\columnwidth, page=3]{2_some_trace_plots.pdf}
	\end{subfigure}

	\caption{Trace-plots of some parameters for the two Markov Chains (left) and posterior density (right) for the simulated data.}
	\label{fig:trace_plots_simulated}
\end{figure}

The assessment of the convergence of the Markov chains is made visually, through their trace plots, as show in Figure \ref{fig:trace_plots_simulated}. Mixing of the chains is appropriate and shows no trend and the estimated densities are smooth. The other parameters perform similarly.

\begin{figure}%
\centering
\includegraphics[width=1\columnwidth]{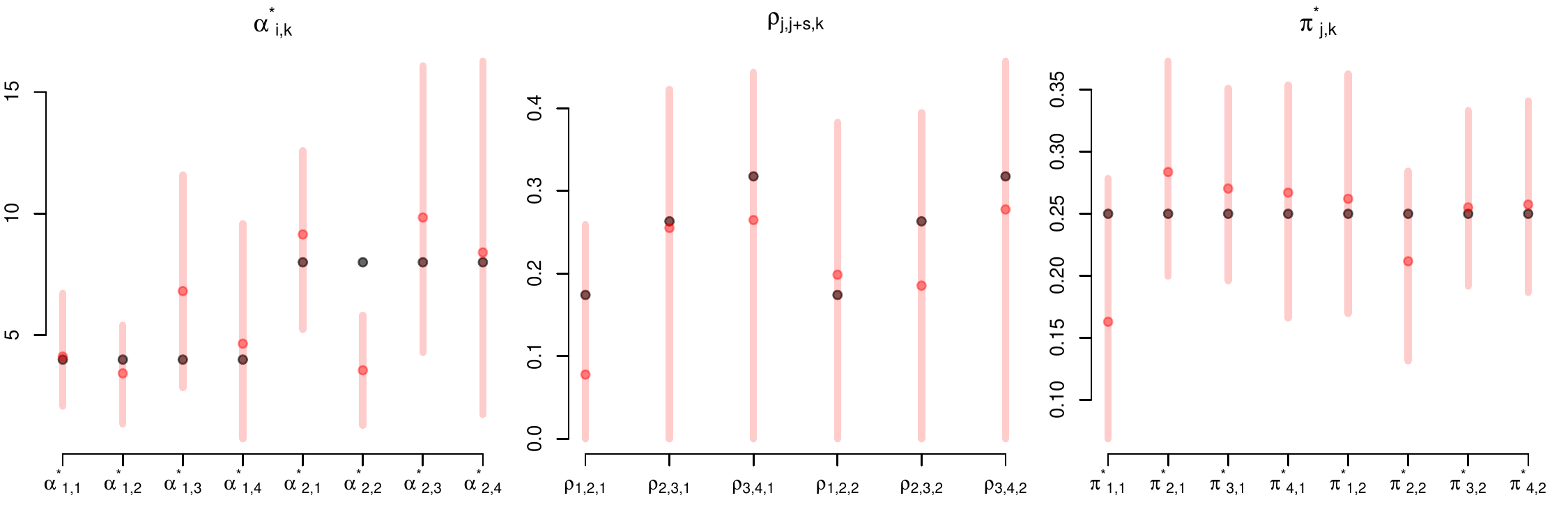}%
\caption{Parameter estimates for simulated data. {\color[rgb]{1,0.8,0.8} $\blacksquare$} 90$\%$ HPD intervals, {\color[rgb]{1,0.48,0.48}$\bullet$} posterior median and {\color[rgb]{0.4,0.4,0.4}$\bullet$} true parameters.}%
\label{fig:HPD}%
\end{figure}

The $90\%$ high posterior density (HPD) intervals using the samples from both chains are shown in Figure \ref{fig:HPD}. The plot is divided in three panels, where each point in the horizontal axis denote one of the unknown (identifiable) parameters: $\alpha^*_{i,k}, \, \rho_{j,j+s,k}$ and $\pi^*_{j,k}$, respectively.

From Figure \ref{fig:HPD} we can see that almost all parameters (apart from $\alpha^*_{2,2}$) lie within their respective 90$\%$ HPD interval (red bar) and most of the point estimates (red dots) are very close to the real values (black dots). The reason why $\alpha_{2,2}^*$ lies away from its HPD interval is because we are using a single replicate of the data. 

\subsection{Real Data}

The data used in this section consists of incremental paid losses for personal auto insurance in the US, a dataset compiled by \cite{meyers2011retrospective}. In order to fit the proposed DGM we only used information up to 1997, which leaves 10 years for testing. Based on the subset of insurers selected by \cite[Appendix A]{meyers2015stochastic}, we selected the 10 largest insurers (by posted reserves in 1997) to perform our analysis on. The group codes for these insures are presented in the first column of Table \ref{tbl:group_code} (in decreasing order of their 1997 reserves). 

\begin{figure}
	\centering
	\begin{subfigure}[b]{0.7\textwidth}
		\includegraphics[width=\columnwidth, page=1]{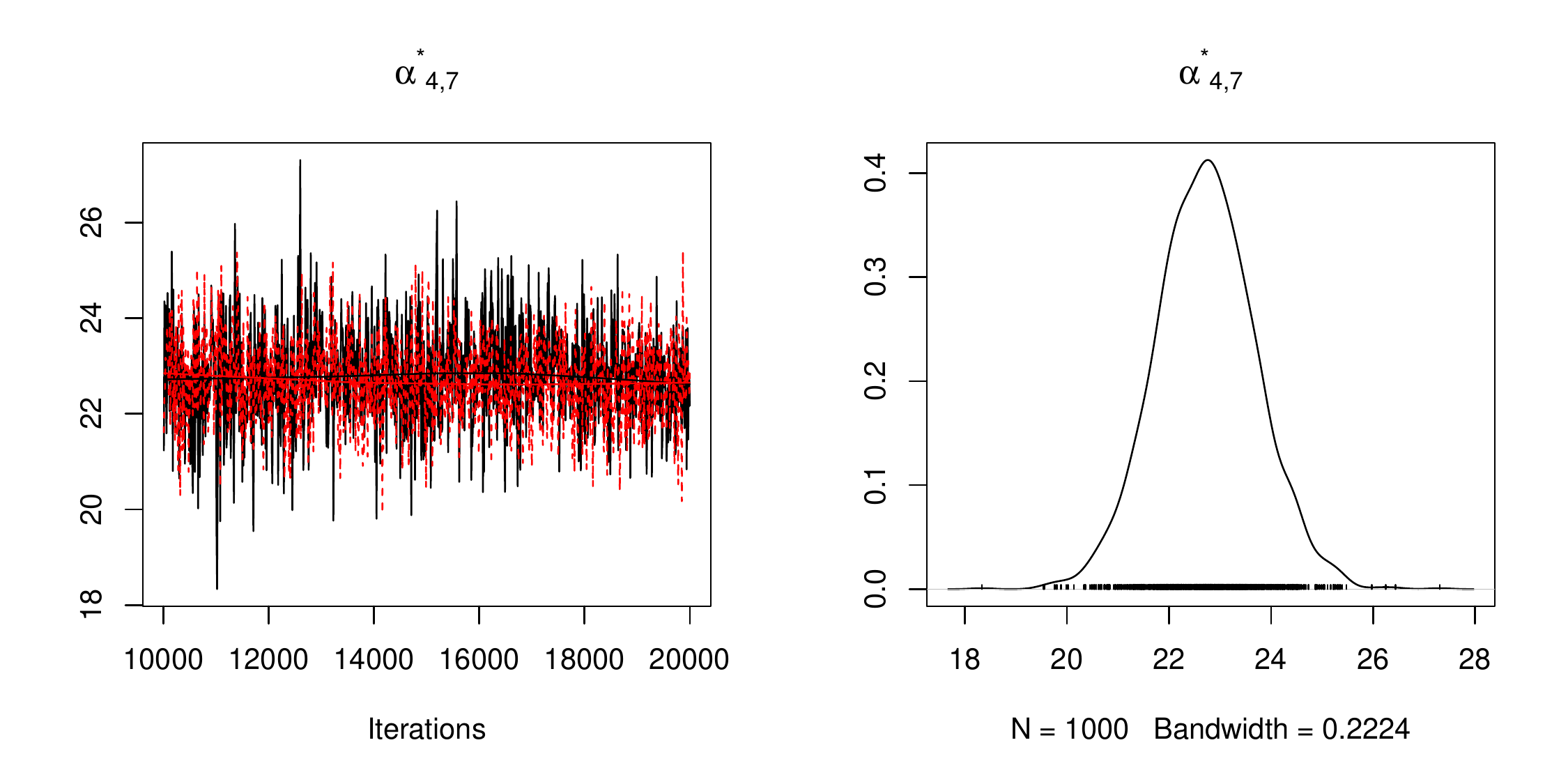}
	\end{subfigure}
	
	\begin{subfigure}[b]{0.7\textwidth}
		\includegraphics[width=\columnwidth, page=2]{3_some_real_trace_plots.pdf}
	\end{subfigure}
	
	\begin{subfigure}[b]{0.7\textwidth}
		\includegraphics[width=\columnwidth, page=3]{3_some_real_trace_plots.pdf}
	\end{subfigure}
	
	\caption{Trace-plots of some parameters for the two Markov Chains (left) and the posterior densities (right) for the real data.}
	\label{fig:trace_plots_real}
\end{figure}

To avoid numerical problems we rescaled the data, first dividing all claims by $1,000$ and then taking its square root. Then played with different specifications of the prior distributions. Specifically we took $p\in\{0,\ldots,5\}$, $a_{\alpha 0} = b_{\alpha 0} \in \{1, \, 10 \}$, $a_{\beta 0} = b_{\beta 0} \in \{1, \, 10\}$, and $a_{\gamma 0} = b_{\gamma 0} = 10$. The combinations of thee values lead to 24 models which are summarised in Table \ref{tbl:prior_choices}. We ran a Gibbs sampler with two parallel chains, each one with 10,000 iterations, a burn-in of 100,000 iterations and keeping one of every 10th iteration to compute posterior estimates. For completeness, some of the trace-plots are presented in Figure \ref{fig:trace_plots_real}, which show a satisfied convergence.

The best model was selected based on two goodness of fit measures: the deviance information criterion (DIC) \citep{spiegelhalter&al:02}, which is a model selection criterion that penalises for model complexity; and the L-measure \citep{ibrahim&laud:94} based on the posterior predictive density and defined as
\[ L(\nu) = \frac{1}{M} \sum_{k=1}^K \sum_{i=2}^n \sum_{j=n-i+2}^n \V(X_{i,j,k}\mid\bx) + \frac{\nu}{M}  \sum_{k=1}^K \sum_{i=2}^n \sum_{j=n-i+2}^n \{\E(X_{i,j,k} \mid\bx) - x_{i,j,k} \}^2, \]
where $\nu\in[0,1]$ is a weighting term which determines a trade-off
between variance and bias, and $M = K n (n-1)/2$ is the number of unknowns.

\begin{figure}%
\centering
\includegraphics[width=\columnwidth]{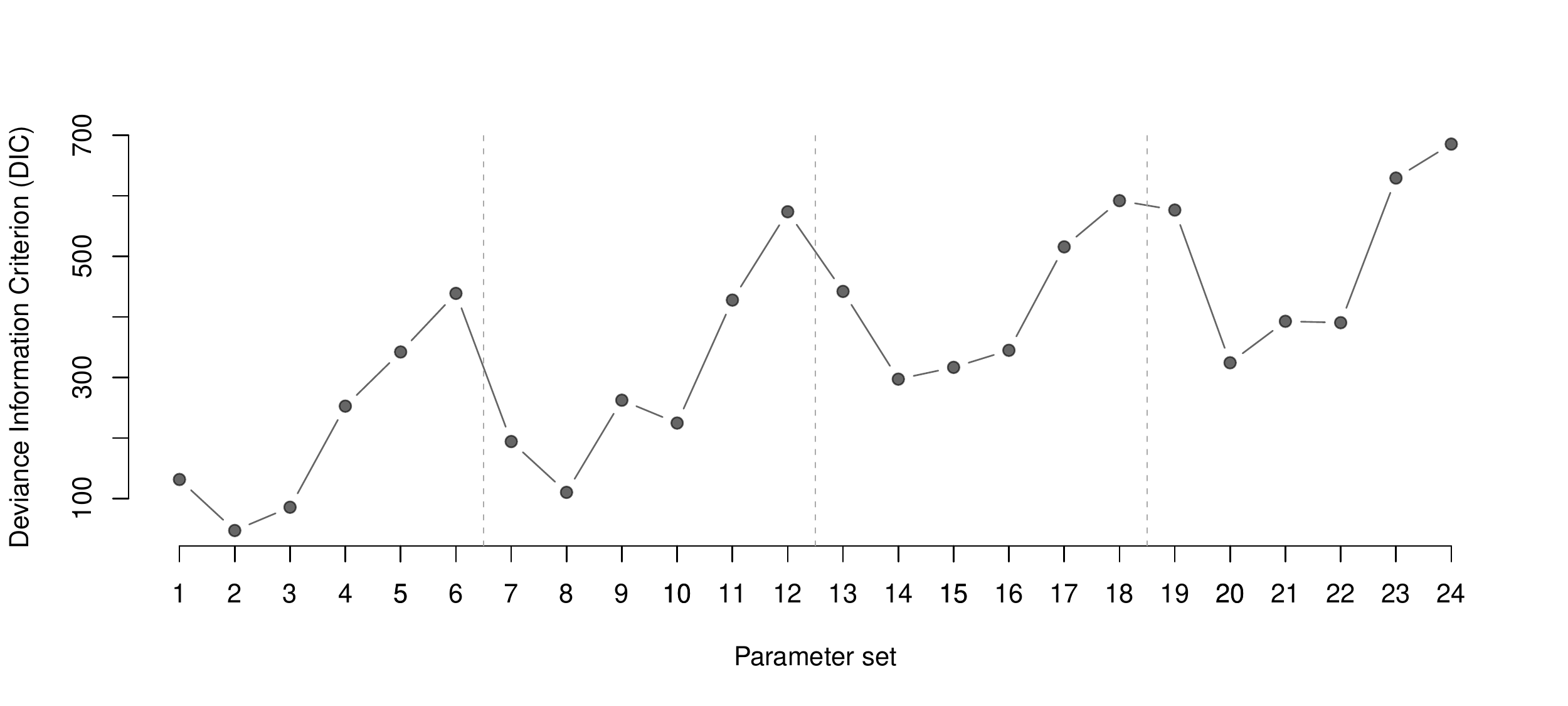}%
\caption{Deviance Information Criterion (DIC) for the 24 models of Table \ref{tbl:prior_choices}. Vertical dotted lines divide the models in four blocks of six models.}%
\label{fig:DIC}%
\end{figure}

Figure \ref{fig:DIC} presents the DIC values obtained for each of the  models. Values are shown in four blocks of 6 models, each block has the same prior specifications but with varying $p$. In the four blocks, the best fitting is obtained with $p=1$, with a shorter difference in the third block for $p=2,3$. Across blocks it seems that there is an increasing trend in the DIC values. Overall the best fitting is achieved by model 2, which corresponds to $p=1$, $a_{\alpha 0} = b_{\alpha 0} = 1$ and $a_{\beta 0} = b_{\beta 0} = 1$. 

\begin{figure}%
\centering
\includegraphics[width=\columnwidth]{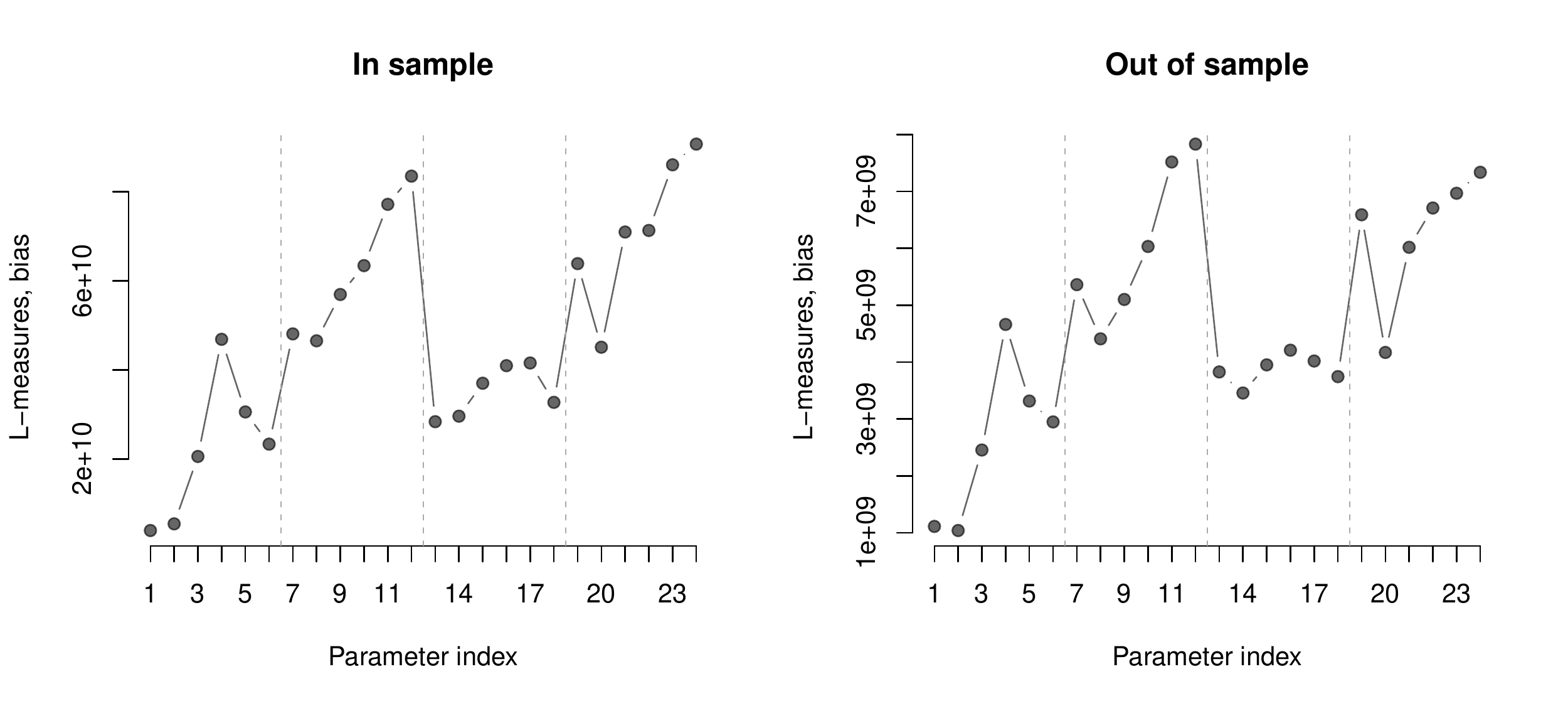}%
\caption{L-measure with $\nu=1/2$ for the 24 models of Table \ref{tbl:prior_choices}.Vertical dotted lines divide the models in four blocks of six models.}%
\label{fig:L_measure}%
\end{figure}

On the other hand, Figure \ref{fig:L_measure} includes the L-measure with $\nu=1/2$ for in-sample data (left panel) and out-of-sample data (right panel). Again results are reported in four blocks of 6 models as in Figure \ref{fig:DIC}. The qualitative behaviour of the measures in- and out-of-sample is very similar and also similar to what is observed with the DIC in Figure \ref{fig:DIC}. However, the increasing trend of the DIC is not shown with the L-measures. The third block has lower values than the second and fourth blocks. Again, the best fitting is achieved by model 2. Therefore, results from this winning model will be discussed in the sequel. 

\begin{figure}%
\centering
\includegraphics[width=\columnwidth, trim =0 11cm 0 0 ]{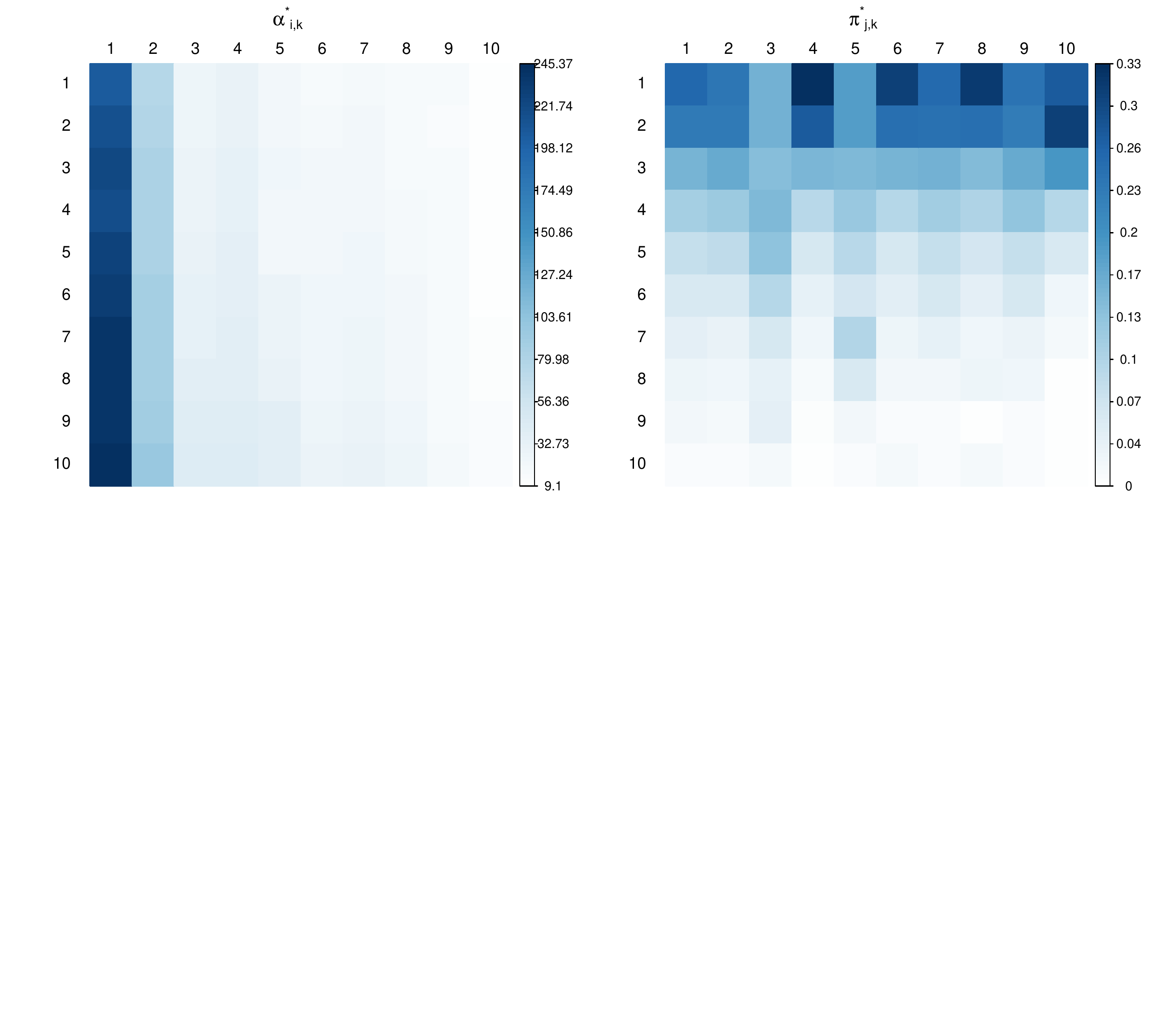}%
\caption{Posterior estimates (medians) for $\alpha_{i,k}^*$ (left) and $\pi_{j,k}^*$ (right).}%
\label{fig:alpha_pi_star}%
\end{figure}

Figure \ref{fig:alpha_pi_star} shows the point estimates, based on the 50\% quantile, for $\alpha_{i,k}^*$ (left panel) and $\pi_{j,k}^*$ (right panel). One can clearly see that the values of $\alpha_{i,k}^*$ (the ultimate total amount for business $k$ with origin at year $i$) decrease along the columns, i.e., when $k$ increases. This is expected, as the business lines are ordered in a decreasing way based on their size (posted reserves in 1997). Although less pronounced, it is also possible to see the \textit{calendar year effect} discussed in \cite{brydon2009calendar}, represented by the fact that $\alpha_{i,k}^*$ is (slightly) increasing in $i$ (across rows). On the other hand, a clear pattern also emerges from the estimates of $\pi_{j,k}^*$, where one can observe a decrease in their values when the development year $j$ increases (across rows). The intuition for this result is that the further we are from the firsts development years $j$, the smaller is the proportion of the ultimate total claim amount $\alpha_{i,k}^*$ expected in development year $j$.

\begin{figure}%
	\centering
	\includegraphics[width=0.8\columnwidth]{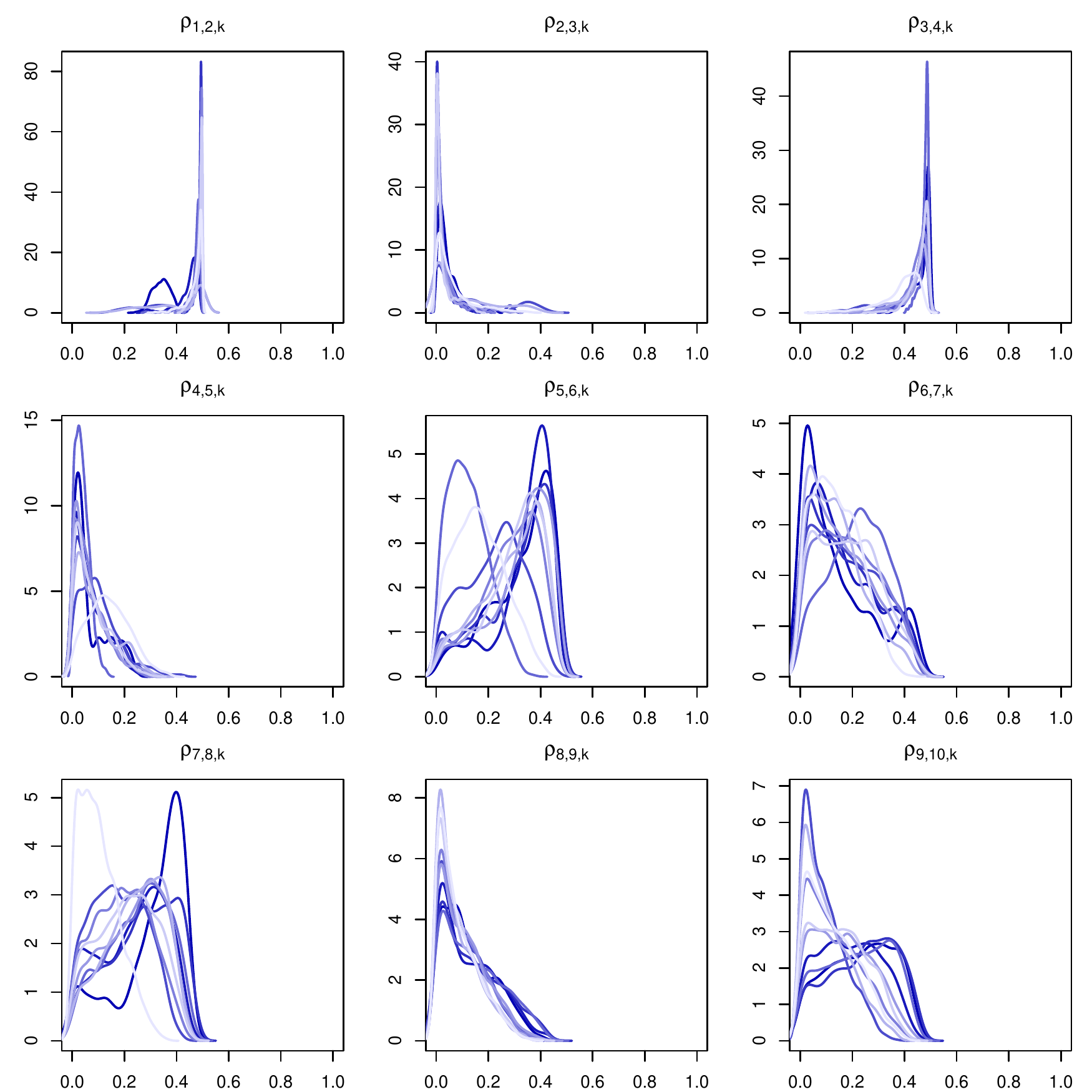}%
	\caption{Posterior densities of the correlation coefficients $\rho_{j,j+1,k}$, $j=1,\ldots,9$. Darker colours denote larger companies while lighter colours denote smaller companies.}
	\label{fig:hist_rho}%
\end{figure}

Since for the best fitting model $p=1$, there are only 9 correlation coefficients $\rho_{j,j+1,k}$, as in \eqref{eq:cor}, for $j=1,\ldots,9$ and for each company $k=1,\ldots,10$. Posterior densities of these coefficients are presented in Figure \ref{fig:hist_rho}. From these plots we can see that, for most development years $j$, all companies have similar correlation distributions. One particularly different case is for development year $j=5$ (center panel), where almost all posterior distributions are left skewed, apart from two, which are right skewed and concentrated on smaller values. A similar situation occurs for development year $j=7$, when only one company's posterior distribution stands out as left skewed. In general, all correlations are likely to be smaller than $0.5$. An interesting fact is that correlations tend to be larger (around 0.5) and with low dispersion for odd years $j=1$, $j=3$ and $j=5$; be very small (less than 0.2) for years $j=2$ and $j=4$; and with a range between 0 and 0.4 for years $j=6,7,8,9$. This could provide a degree of importance for each development year in the whole triangle. 

\begin{figure}%
	\includegraphics[width=\columnwidth, page=1]{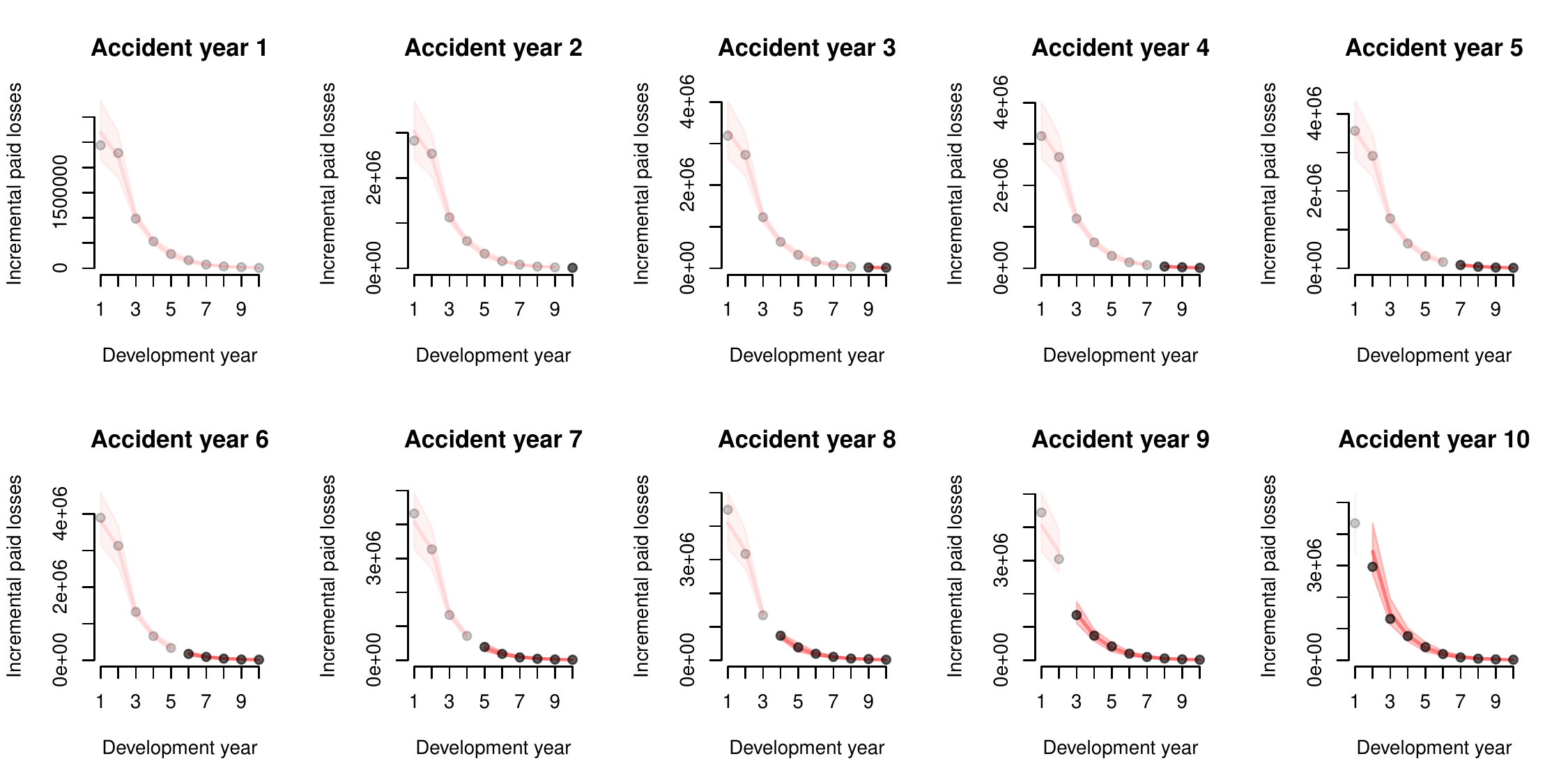}%
	\caption{Largest company in the sample (code 1767). Posterior predictions of incremental paid losses per development year $i$, for all accident years $j$. {\color[rgb]{0.8,0.8,0.8}$\bullet$} Observed claims (upper triangle), {\color[rgb]{0.4,0.4,0.4}$\bullet$} Non-observed claims (lower triangle), {\color[rgb]{1,0.8,0.8} $\blacksquare$} 95\% credibility interval, {\color[rgb]{1,0.48,0.48} $-$} median. }%
	\label{fig:posterior_predictive_1}%
\end{figure}

\begin{figure}%
	\includegraphics[width=\columnwidth, page=10]{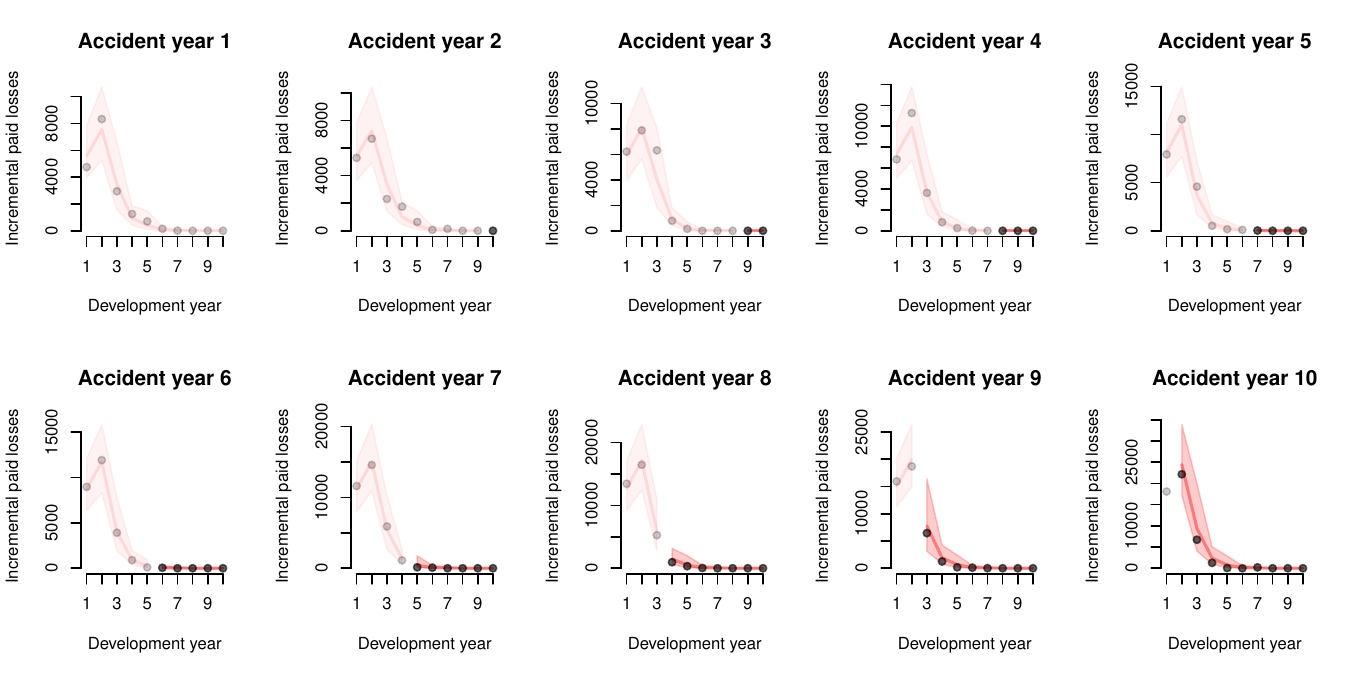}%
	\caption{Smallest company in the sample (code 692). Posterior predictions of incremental paid losses per development year $i$, for all accident years $j$. {\color[rgb]{0.8,0.8,0.8}$\bullet$} Observed claims (upper triangle), {\color[rgb]{0.4,0.4,0.4}$\bullet$} non-observed claims (lower triangle), {\color[rgb]{1,0.8,0.8} $\blacksquare$} 95\% credibility interval, {\color[rgb]{1,0.48,0.48} $-$} median. }%
	\label{fig:posterior_predictive_2}%
\end{figure}

We compute posterior predictive distributions for $X_{i,j,k}$ in the lower-right triangle, i.e. for $i=2,\ldots,n$, $j=n+2-i,\ldots,n$ and $n=10$. We present the median (dark red) together with the 95\% credibility intervals (light red), for two companies: $k=1$, which corresponds to the largest company, State Farm Mut Grp (code 1767), in Figure \ref{fig:posterior_predictive_1}; and $k=10$, which corresponds to the smallest company, Wawanesa Ins Grp (code 692), in Figure \ref{fig:posterior_predictive_2}. For each accident year the light grey dots represent the observed loss data (upper triangle) and the dark grey dots are observations not provided to the model (lower triangle). 

For the largest company (Figure \ref{fig:posterior_predictive_1}), the model is able to fit the in-sample data perfectly well, and it produces  very precise predictions. For the smallest company (Figure \ref{fig:posterior_predictive_2}) the model also performs well, even being able to fit the hump at development year 2 in all accident years. The credible intervals are, perhaps, a little wider than those for the largest company ($k=1$). 

To see the advantage of our modeling with respect to standard methods, we provide a comparison between the proposed DGM, which considers dependence across development years and pulls strength across the ten companies, with the over-dispersed Poisson (ODP). We fitted ten independent ODP models using the \texttt{R}-package \texttt{ChainLadder} \citep{ChainLadder}. For the comparison we show the reserves per accident year $i$ as well as the aggregated reserves for the ten companies.

\begin{figure}
\centering
\begin{subfigure}{.5\textwidth}
  \centering
  \includegraphics[width=\columnwidth, page=1]{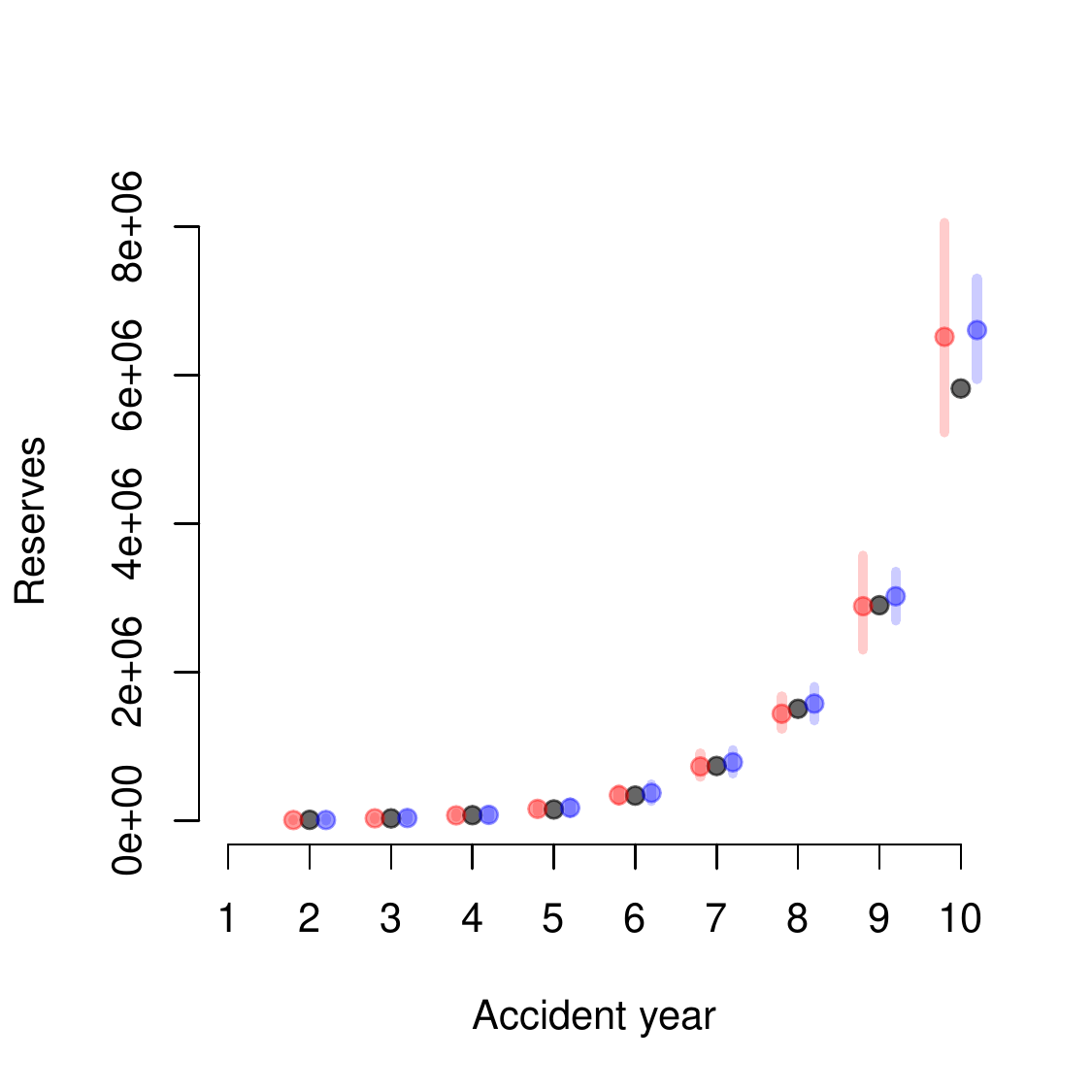}
  \label{fig:sub1}
\end{subfigure}%
\begin{subfigure}{.5\textwidth}
  \centering
  \includegraphics[width=\columnwidth, page=10]{4_Reserves.pdf}
  \label{fig:sub2}
\end{subfigure}
\caption{Reserves for each accident year. (Left) largest company (code 1767) and (right) smallest company (code 692). {\color[rgb]{1,0.6,0.6}$\bullet$} median and 95\% CI for DGM, {\color[rgb]{0.42,0.42,0.42}$\bullet$} true claims and {\color[rgb]{0.6,0.6,1}$\bullet$} median and 95\% CI for ODP.}%
\label{fig:reserves}%
\end{figure}

Figure \ref{fig:reserves} presents the reserves estimates from the DGM (red) and the ODP model (blue) for each accident year $i=1,\ldots,10$, for the largest company $k=1$ (code 1767) and the smallest company $k=10$ (code 692). For the DGM, posterior predictive estimates are given by the median and the quantiles to form a 95\% CI. For the ODP, predictions are based on maximum likelihood estimates and bootstrap 95\% CI. 
For comparison we also include the true observed claims (grey). For both companies, the first seven accident years ($i=1,\ldots,7$) both estimates, DGM and ODP are very precise, but for the last three accident years ($i=8,9,10$) interval estimates are wider with the DGM showing more dispersion than the ODP, however, for these last thee years, DGM point estimates show less bias than those from ODP model.

\begin{figure}%
	\centering
	\includegraphics[width=0.5\columnwidth]{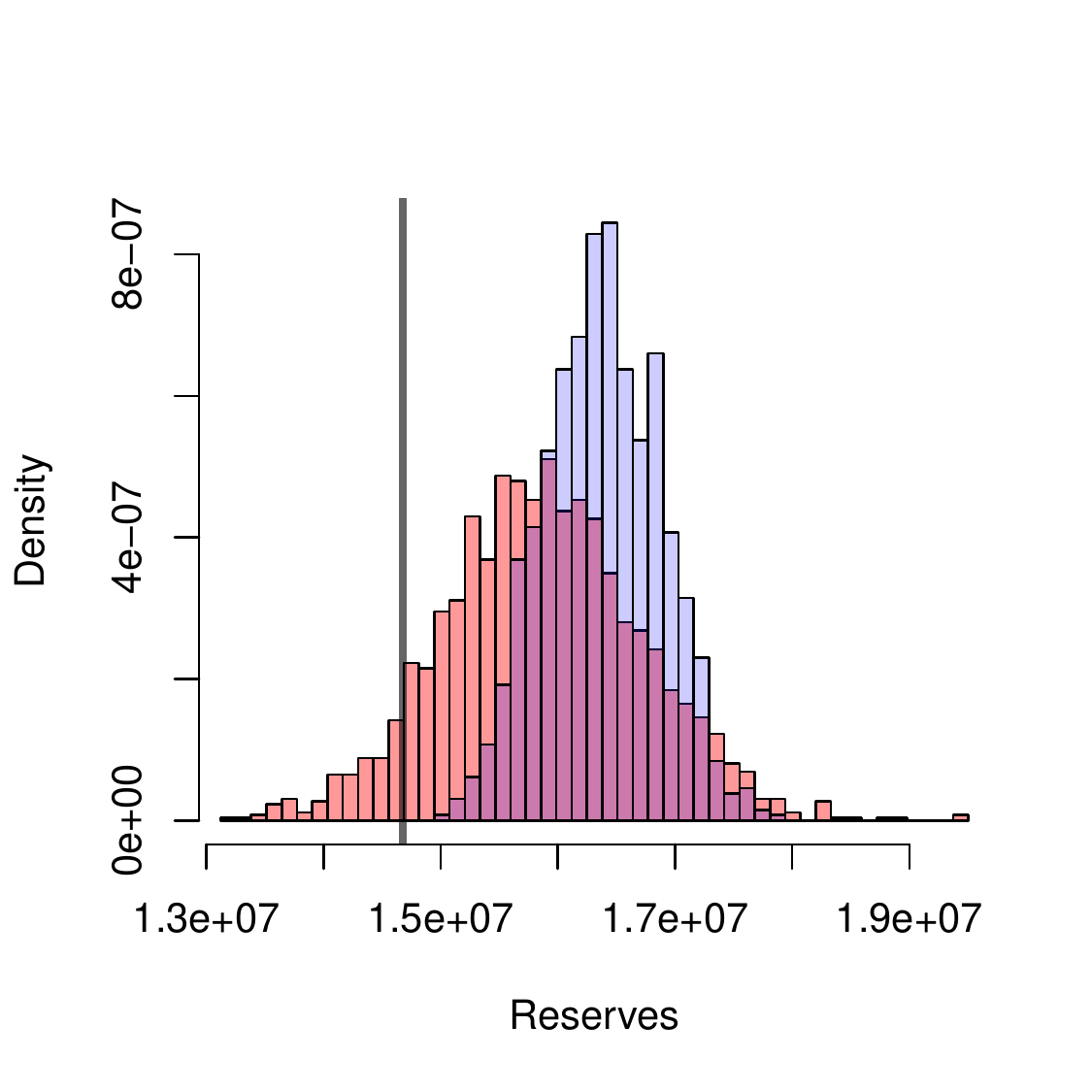}%
	\caption{Predictive distributions of the aggregated reserves for all ten companies. {\color[rgb]{0.42,0.42,0.42}$\textbf{--}$} True claims; {\color[rgb]{1,0.8,0.8} $\blacksquare$} DGM; {\color[rgb]{0.8,0.8,1}$\blacksquare$} ODP (Bootstrap). }%
	\label{fig:hist_reserve}%
\end{figure}

To further analyse the prediction performance, we aggregate the reserves for all ten companies and compute the predictive distribution of the aggregated reserve, which is shown in Figure \ref{fig:hist_reserve} as a probability histogram. As postulated, the DGM (orange) has larger variance and smaller bias than the ODP (blue). The pink area corresponds to the intersection of both histograms. Additionally, we can see that the predictive distribution of the ODP lies away from the true observed reserve (grey vertical line), whereas the predictive distribution of the DGM captures well the true reserve. 

It is important to notice that this exercise was performed to showcase the ability of the proposed dependent Gamma model to perform accurate forecasts of future claims payments. For this task the point estimates were chosen as the 50$\%$ quantile (median), and the credibility intervals are such that the posterior probability of the median being in these sets is 95$\%$. In practice, modern solvency regulations, such as Solvency II \citep{solvencyii} and the Swiss Solvency Test \citep{SST_technical_document} requires reserves to be computed in a much more conservative fashion. For example, the first requires reserves to be computed based on the $99.5\%$-quantile or value at risk (VaR) of the distribution of the losses, while the latter uses the Expected Shortfall (also called conditional VaR) at the level $99\%$ of the same distribution. 

Both quantities, and their related credibility intervals, can be easily computed from the model's outputs, as the Markov Chain Monte Carlo algorithm returns samples of the required posterior distribution.

\section{Concluding remarks}

We presented an easily interpretable stochastic model for claims reserving (the DGM -- dependet gamma model) that captures dependence across development years within a single triangle and combines information from multiple triangles through a judicious choice of prior distributions. Dependence is of any order $p\geq0$ of past developments years, with a very appealing parametrisation that can be easily interpreted. 

Posterior inference of our model can easily be obtained through a MCMC procedure implemented in \texttt{rjags}. Code is available upon request from the second author. 

Our examples, using the NAIC data set, show that our method works well for both small and large companies, with more accurate predictions than the benchmark model (over-dispersed Poisson). Moreover, dependence is summarised through correlation coefficients across different development  years within the same company ($\rho_{j,j+s,k}$) and the other interpretable parameters ($\alpha_{i,k}^*$ and $\pi_{j,k}^*$) provide some insight of the ultimate total claims per accident year $i$ and how it is divided into the development years $j$, respectively for each triangle $k$. 

Due to the simple hierarchical structure of the DGM, it may be possible (at least numerically) to study further quantities, such as the claims development result (CDR) -- the difference between a) the reserves predicted at time $t$ and b) the reserves predicted at time $t+1$ plus the payments made at time $t+1$. For more information on this ``one-year risk'', see \cite{ohlsson2009one} or \cite{merz2015claims}.

Our DGM construction \eqref{eq:mod} is very flexible and can also be adapted to seasonal dependence. That is, if the seasonality of the data is $s$, then the model would be 
$X_{i,j,k}\mid\bZ\sim\ga(\alpha_{i,k}+\sum_{l=0}^p Z_{i,j-sl,k},\,\beta_{j,k}+\sum_{l=0}^p \gamma_{j-sl,k})$. For instance, if the data were available in a monthly basis, we can assume a seasonality of order $s=12$ so that the incremental claims made in January of the current year can depend on the January claims of the previous $p$ years. This and other generalisations are worth studying in the future.

\section*{Acknowledgements}
This research was supported by the School of Applied Mathematics (EMAp) from Funda\c{c}\~ao Getulio Vargas (FGV), Brazil. The first author acknowledges support from \textit{Asociaci\'on Mexicana de Cultura A.C.}--Mexico.

\section*{Appendix}

Posterior conditional distributions of model parameters $\btheta$ and the hyper-parameters, for $i,j=1,\ldots,n$, $k=1,\ldots,K$, are:
\begin{enumerate}
\item[(i)] Full conditional of $\alpha_{i,k}$
$$f(\alpha_{i,k}\mid\rest)\propto
\left\{e^{-(b_{\alpha i}+\sum_{j=1}^{n+1-i}\gamma_{j,k})}\prod_{j=1}^{n+i-1}(\beta_{j,k}+\sum_{l=0}^p\gamma_{j-l,k})x_{i,j,k}\right\}^{\alpha_{k,i}}\alpha_{i,k}^{a_{\alpha i}+\sum_{j=1}^{n+1-i}z_{i,j,k}-1}$$
$$\hspace{-2.5cm}\times\frac{1}{\prod_{j=1}^{n+1-i}\Gamma(\alpha_{i,k}+\sum_{l=0}^{p}z_{i,j-l,k})}I(\alpha_{i,k}>0)$$
\item[(ii)] Full conditional of $\beta_{j,k}$
$$f(\beta_{j,k}\mid\rest)\propto\left(\beta_{j,k}+\sum_{l=0}^p\gamma_{j-l,k}\right)^{\sum_{i=1}^{n+1-j}(\alpha_{i,k}+\sum_{l=0}^pz_{i,j-l,k})}e^{-\beta_{j,k}(b_{\beta j}+\sum_{i=1}^{n+1-j}x_{i,j,k})}$$ 
$$\hspace{-5cm}\times\beta_{j,k}^{a_{\beta j}-1}I(\beta_{j,k}>0)$$
\item[(iii)] Full conditional of $\gamma_{j,k}$
$$\hspace{-4cm}f(\gamma_{j,k}\mid\rest)\propto\left\{\prod_{r=0}^p\left(\beta_{j+r,k}+\sum_{l=0}^p\gamma_{j+r-l,k}\right)^{\sum_{i=1}^{n+1-j}(\alpha_{i,k}+\sum_{l=0}^pz_{i,j+r-l,k})}\right\}$$ 
$$\hspace{1.5cm}\times e^{-\gamma_{j,k}\left(b_{\gamma j}+\sum_{r=0}^p\sum_{i=1}^{n+1-j}x_{i,j+r,k}+\sum_{i=1}^{n+1-j}\alpha_{i,k}\right)}\,\gamma_{j,k}^{a_{\gamma j}+\sum_{i=1}^{n+1-j}z_{i,j,k}-1}I(\gamma_{j,k}>0)$$ 
\item[(iv)] Full conditional of $z_{i,j,k}$
$$\hspace{-3cm}f(z_{i,j,k}\mid\rest)\propto\left\{\alpha_{i,k}\gamma_{j,k}\prod_{r=0}^{p}(\beta_{j+r,k}+\sum_{l=0}^p\gamma_{j+r-l,k})x_{i,j+r,k}\right\}^{z_{i,j,k}}$$ 
$$\hspace{2cm}\times\Gamma^{-1}(z_{i,j,k}+1)\left\{\prod_{r=0}^p\Gamma^{-1}(\alpha_{i,k}+\sum_{l=0}^pz_{i,j+r-l,k})\right\}I_{\{0,1,\ldots\}}(z_{i,j,k})$$
\item[(v)] Full conditionals of $a_{\alpha i}$ and $b_{\alpha i}$
$$f(a_{\alpha i}\mid\rest)\propto \frac{(bv)^{Ka_{\alpha i}}}{\Gamma^K(a_{\alpha i})}\left(\prod_{k=1}^K\alpha_{i,k}\right)^{a_{\alpha i}}\ga(a_{\alpha i}\mid a_{\alpha 0},b_{\alpha 0})\quad\mbox{and}$$ $$f(b_{\alpha i}\mid\rest)\propto (b_{\alpha i})^{Ka_{\alpha i}}e^{-b_{\alpha i}\sum_{k=1}^K\alpha_{i,k}}\ga(b_{\alpha i}\mid a_{\alpha 0},b_{\alpha 0})$$
\item[(vi)] Full conditionals of $a_{\beta j}$ and $b_{\beta j}$
$$f(a_{\beta j}\mid\rest)\propto \frac{(b_{\beta j})^{Ka_{\beta j}}}{\Gamma^K(a_{\beta j})}\left(\prod_{k=1}^K\beta_{j,k}\right)^{a_{\beta j}}\ga(a_{\beta j}\mid a_{\beta 0},b_{\beta 0})\quad\mbox{and}$$ $$f(b_{\beta j}\mid\rest)\propto (b_{\beta j})^{Ka_{\beta j}}e^{-b_{\beta j}\sum_{k=1}^K\beta_{j,k}}\ga(b_{\beta j}\mid a_{\beta 0},b_{\beta 0})$$
\item[(vii)] Full conditionals of $a_{\gamma j}$ and $b_{\gamma j}$
$$f(a_{\gamma j}\mid\rest)\propto \frac{(b_{\gamma j})^{Ka_{\gamma j}}}{\Gamma^K(a_{\gamma j})}\left(\prod_{k=1}^K\gamma_{j,k}\right)^{a_{\gamma j}}\ga(a_{\gamma j}\mid a_{\gamma 0},b_{\gamma 0})\quad\mbox{and}$$ $$f(b_{\gamma j}\mid\rest)\propto (b_{\gamma j})^{Ka_{\gamma j}}e^{-b_{\gamma j}\sum_{k=1}^K\gamma_{j,k}}\ga(b_{\gamma j}\mid a_{\gamma 0},b_{\gamma 0})$$
\end{enumerate}

\newpage
\bibliographystyle{authordate1}
\bibliography{../bibliography}

\newpage

\begin{table}
$$\begin{array}{|c|ccccccc|} 
\hline
\multicolumn{8}{|c|}{\text{Business }k} \\
\hline
\text{Year of} & \multicolumn{7}{c|}{\text{Development year}} \\ 
\text{origin} & 1 & 2 & \cdots & j & \cdots & n-1 & n \\ \hline
1 & X_{1,1,k} & X_{1,2,k} & \cdots & X_{1,j,k} &  & X_{1,n-1,k} & X_{1,n,k} \\ 
2 & X_{2,1,k} & X_{2,2,k} & \cdots & X_{2,j,k} &  & X_{2,n-1,k} &  \\ 
\vdots & \vdots & \vdots & \cdots & \vdots &  &  &  \\ 
i & X_{i,1,k} & X_{i,2,k} & \cdots & X_{i,n+1-i,k} &  &  &  \\ 
\vdots & \vdots & \vdots &  &  &  &  &  \\ 
n-1 & X_{n-1,1,k} & X_{n-1,2,k} &  &  &  &  &  \\ 
n & X_{n,1,k} &  &  &  &  &  &  \\ \hline
\end{array}$$
\caption{{\protect\small Run-off triangle of available data for business $k=1,\ldots,K$.}}
\label{tab:rotk}
\end{table}

\begin{table}[ht]
\centering
\begin{tabular}{lll}
  \toprule
$k$	& \textbf{Group codes}	&	\textbf{Group Name}	\\
  \midrule	
1	&	1767	&	State Farm Mut Grp	\\
2	&	2003	&	United Services Automobile Asn Grp	\\
3	&	7080	&	New Jersey Manufacturers Grp	\\
4	&	4839	&	FL Farm Bureau Grp	\\
5	&	388	&	Federal Ins Co Grp	\\
6	&	1090	&	Kentucky Farm Bureau Mut Ins Grp	\\
7	&	3240	&	NC Farm Bureau Ins Grp	\\
8	&	6947	&	Tenn Farmers Mut	\\
9	&	620	&	Employers Mut Co Of Des Moines	\\
10	&	692	&	Wawanesa Ins Grp	\\ 
   \bottomrule
\end{tabular}
\caption{Group codes and names for the ten companies analysed. These correspond to the largest companies based on their posted reserves in 1997.}
\label{tbl:group_code}
\end{table}

\begin{table}[ht]
	\centering
	\begin{tabular}{cccc}
		\toprule
\textbf{Model}	&	\textbf{p}	&	$a_{\alpha 0} = b_{\alpha 0}$	&	$a_{\beta 0} = b_{\beta 0}$	\\
		\midrule	
1	&	0	&	1	&	1	\\
2	&	1	&	1	&	1	\\
$\vdots$ & $\vdots$ & $\vdots$ & $\vdots$ \\
6	&	5	&	1	&	1	\\
		\midrule	
7	&	0	&	10	&	1	\\
$\vdots$ & $\vdots$ & $\vdots$ & $\vdots$ \\
12	&	5	&	10	&	1	\\
		\midrule	
13	&	0	&	1	&	10	\\
$\vdots$ & $\vdots$ & $\vdots$ & $\vdots$ \\
18	&	5	&	1	&	10	\\
\midrule
19	&	0	&	10	&	10	\\
$\vdots$ & $\vdots$ & $\vdots$ & $\vdots$ \\
24	&	5	&	10	&	10	\\
		\bottomrule
	\end{tabular}
	\caption{Prior specification for the 24 models. The table is sorted (in ascending order) by $p$, then $a_{\alpha 0}$ and finally by $a_{\beta 0}$.}
	\label{tbl:prior_choices}
\end{table}

\newpage

\end{document}